\documentclass[a4paper,fleqn,usenatbib]{mnras2}
\usepackage{newtxtext,newtxmath}
\usepackage[T1]{fontenc}
\usepackage{ae,aecompl}
\usepackage{hyperref}
\usepackage{graphicx}	
\usepackage{amsmath}	
\usepackage{amssymb}	
\usepackage{natbib}
\usepackage{color}
\usepackage{ulem}
\usepackage{mathtools}
\usepackage{threeparttable}
\usepackage{booktabs}
\usepackage{verbatim}
\usepackage{hyperref}

\title[SNe Ia in NS+He star systems]{Type Ia supernovae in NS+He star systems and the isolated mildly recycled pulsars}

\author[Y. Guo et al.]{
	Yun-Lang Guo,$^{\rm 1,2,3}$\thanks{E-mail:yunlang@ynao.ac.cn}
	Bo Wang,$^{\rm 1,2,3}$\thanks{E-mail:wangbo@ynao.ac.cn}
	Cheng-Yuan Wu,$^{\rm 1,2,3}$
	Wen-Cong Chen,$^{\rm 4}$
	Long Jiang$^{\rm 4}$
	and
	Zhan-Wen Han$^{\rm 1,2,3}$
	\\
$^{1}$Yunnan Observatories, Chinese Academy of Sciences, Kunming 650216, China\\
$^{2}$International Centre of Supernovae, Yunnan Key Laboratory, Kunming 650216, China\\
$^{3}$University of Chinese Academy of Sciences, Beijing 100049, China\\
$^{4}$School of Science, Qingdao University of Technology, Qingdao 266525, China\\
}

\date{Accepted XXX. Received YYY; in original form ZZZ}

\pubyear{2022}

\begin{document}
\label{firstpage}
\pagerange{\pageref{firstpage}--\pageref{lastpage}}
\maketitle

\begin{abstract}
Type Ia supernovae (SNe Ia) are successful cosmological distance indicators
and important element factories in the chemical evolution of galaxies.
They are generally thought to originate from thermonuclear explosions of
carbon-oxygen white dwarfs in close binaries.
However,
the observed diversity among SNe Ia implies that
they have different progenitor models.
In this article, we performed the long-term evolution of NS+He star binaries
with different initial He star masses
($M_{\rm He}^{\rm i}$)
and orbital periods ($P_{\rm orb}^{\rm i}$) for the first time,
in which the He star companions can explode as SNe Ia eventually.
Our simulations indicate that
after the He stars develop highly degenerate oxygen-neon (ONe) cores
with masses near the Chandrasekhar limit,
explosive oxygen burning can be triggered due to the
convective Urca process.
According to these calculations,
we obtained an initial parameter space for the production of SNe Ia
in the $\rm log\,$$P^{\rm i}_{\rm orb}-M^{\rm i}_{\rm He}$ plane.
Meanwhile,
we found that isolated mildly recycled pulsars can be formed after
He stars explode as SNe Ia in NS+He star binaries,
in which the isolated pulsars have minimum spin periods
($P_{\rm spin}^{\rm min}$) of $\sim 30-110$\,ms
and final orbital velocities of $\sim \rm 60-360\,km\,s^{-1}$,
corresponding to initial orbital periods of $0.07-10$\,d.
Our work suggests that the NS+He star channel may contribute
to the formation of isolated mildly recycled pulsars with
velocity $\rm \lesssim 360\,km\,s^{-1}$ in observations,
and such isolated pulsars should locate in the region
of pulsars with massive WD companions
in the $P_{\rm spin}-\dot P_{\rm spin}$ diagram.
\end{abstract}


\section{Introduction}

Type Ia supernovae (SNe Ia) are excellent cosmological distance indicators
due to their uniform luminosities,
revealing the accelerating expansion of the Universe driven by dark energy 
\citep[e.g.][]{1998AJ....116.1009R, 1999ApJ...517..565P, 2011NatCo...2..350H}.
They also play an important role in the chemical evolution of galaxies
owing to their contributions to the abundance of heavy elements
\citep[e.g.][]{1983A&A...118..217G, 1986A&A...154..279M}.
Although SNe Ia are important objects for cosmology and chemical evolution of galaxies,
their progenitors remain controversial
\citep[e.g.][]{2014ARA&A..52..107M, 2018RAA....18...49W, 2023RAA....23h2001L}.

It is generally believed that SNe Ia are produced by
thermonuclear explosions of carbon-oxygen white dwarfs (CO WDs),
in which the WDs have masses close to the Chandrasekhar limit
($M_{\rm Ch}$; e.g. \citealt{1960ApJ...132..565H, 1997Sci...276.1378N}).
Two widely studied progenitor models have been proposed to explain the formation of SNe Ia,
i.e. the single-degenerate (SD) model and the double-degenerate (DD) model
\citep[e.g.][]{1973ApJ...186.1007W, 1984ApJ...286..644N, 1984ApJS...54..335I}.
In the SD model, a CO WD accretes H/He-rich material from a main-sequence star, a red giant star
or a He star until thermonuclear explosions occur
\citep[e.g.][]{1996ApJ...470L..97H, 1997A&A...322L...9L, 2000A&A...362.1046L, 2004MNRAS.350.1301H, 2006MNRAS.368.1095H, 2009MNRAS.395..847W,2017MNRAS.469.4763M, 2019ApJ...871...31A, 2020MNRAS.495.1445W}.
In the DD model,
two WDs merge in a close binary caused by gravitational wave radiation,
and SNe Ia may be produced if their total mass exceeds $M_{\rm Ch}$
\citep[e.g.][]{1984ApJ...277..355W, 2001A&A...365..491N, 2009ApJ...699.2026R,
2012ApJ...755L...9C, 2018MNRAS.473.5352L, 2019MNRAS.483..263W}.
Aside from the SD and DD model,
some alternative models have been proposed to explain
the observed diversity among SNe Ia,
such as the hybrid CONe WD model, the core-degenerate model, and the double WD collision model, etc
\citep[e.g.][]{2009MNRAS.399L.156R, 2011MNRAS.417.1466K,2013MNRAS.431.1541S, 2015MNRAS.447.2696D}.

\citet{2006PhDT.......214W} studied the formation of SNe Ia through the evolution of isolated He stars.
Their simulations indicate that
He stars can develop highly degenerate oxygen-neon (ONe) cores,
and then explosive oxygen burning may be triggered due to
the ignition of central residual carbon
\citep[see also][]{2008ASPC..391..359W}.
Following the work of \citet{2006PhDT.......214W},
\citet{Antoniadis2020A&A} recently found that explosive oxygen burning
can be triggered at central densities of
$\sim 1.8-5.9\times10^9\rm \,g\,cm^{-3}$
\citep[see also][]{2022A&A...668A.106C}.
In order to explore the importance of central residual carbon,
\citet{Antoniadis2020A&A} artificially set the reaction rate of carbon burning to be zero
after main carbon burning phase.
As a result,
$e$-capture on $^{20}\rm Ne$ occurs,
and then oxygen is ignited at central density $\gtrsim 10^{10}\rm \,g\,cm^{-3}$,
resulting in the formation of electron-capture SNe (EC-SNe) instead of SNe Ia.
In the traditional picture, 
$e$-capture on $^{24}\rm Mg$ and $^{20}\rm Ne$ can take place
for ONe cores with masses close to $M_{\rm Ch}$,
and then the ONe cores eventually collapse into neutron stars
\citep[NSs; e.g.][]{Nomoto1984, 1987ApJ...322..206N, 2013ApJ...771...28T, 2013ApJ...772..150J, 2017PASA...34...56D, 2022MNRAS.513.4802A}.
However,
recent studies indicate that the ONe cores with masses close to $M_{\rm Ch}$
may explode as SNe Ia due to the ignition of central residual carbon
\citep[e.g.][]{2006PhDT.......214W,Antoniadis2020A&A,2022A&A...668A.106C}.

It is well known that the evolutionary track of stars in close binaries is
quite different from that of single stars, and therefore the evolutionary outcomes
\citep[see, e.g.][]{2002MNRAS.329..897H}.
For example, the mass-transfer process will change the mass-loss rate of donor,
resulting in different final core masses \citep[e.g.][]{2012ARA&A..50..107L, 2015MNRAS.451.2123T}.
Meanwhile, it has been suggested that most massive stars are in close binaries
and interact with their companion star
\citep[e.g.][]{2012Sci...337..444S,2012MNRAS.424.1925C}.
Especially,
isolated pulsars may be left behind
if He stars explode as SNe Ia in the NS+He star systems.

Accordingly, the purpose of this article is to explore the SNe Ia
that occur in NS+He star systems,
in which the He stars experience explosive oxygen burning.
This article is organized as follows.
In Sect. 2, we introduce the adopted basic assumptions and methods for numerical calculations. 
In Sect. 3, we give the calculated results,
including the evolution from He stars to SNe Ia
and the initial parameter space for SNe Ia.
In Sect. 4,
we show the final evolutionary outcomes of NS+He star binaries
after the He star companions explode as SNe Ia
(i.e. isolated mildly recycled pulsars),
and the comparison between simulations and observations.
Finally, we present relevant discussions in Sect. 5 and a summary in Sect. 6.
\section{Numerical methods and assumptions}
Employing the stellar evolution code Modules for Experiments in Stellar Astrophysics
(MESA, version 10398; see \citealt{2011ApJS..192....3P, 2013ApJS..208....4P, 2015ApJS..220...15P, 2018ApJS..234...34P}),
we calculated the detailed binary evolution of NS+He star systems,
in which the initial NS mass ($M_{\rm NS}^{\rm i}$) is set to be $1.35\,\rm M_\odot$.
In our simulations,
the NS is assumed to be a point mass.
We built a grid of binary models for a typical Population I with metallicity $Z=0.02$,
and the composition of zero-age main-sequence (ZAMS) He stars is $98\%$ helium and $2\%$ metallicity
\citep[e.g.][]{Antoniadis2020A&A, 2021MNRAS.506.4654W}.
During the mass-transfer phase,
we adopted the `kolb' scheme to compute the mass-transfer rate
\citep[][]{kolb1990A&A}.

For the helium accretion in NS binaries,
we adopted a model with $\alpha=0$, $\beta=0.5$ and $\delta=0$
\citep[][]{2006csxs.book..623T,2016ApJ...830..131C,2021MNRAS.506.4654W},
in which $\alpha$, $\beta$ and $\delta$ are the fractions of the mass lost
from the vicinity of the donor star,
the fraction of mass lost from the vicinity of the NS,
and the fraction of mass lost from circumbinary co-planar toroid, respectively.
Accordingly,
the mass-accretion rate for NS can be defined as $\dot M_{\rm acc}=(1-\beta)\dot M_{\rm tran}$,
where $\dot M_{\rm tran}$ is the mass-transfer rate.
We set the Eddington accretion
rate ($\dot M_{\rm Edd}$) to be $3 \times 10^{-8}\,\rm M_\odot \rm yr^{-1}$
\citep[e.g.][]{2002MNRAS.331.1027D,chen2011A&A}.
We assume that the unprocessed matter is ejected from the vicinity of the NS
if mass-accretion rate is larger than the Eddington accretion rate
\citep[e.g.][]{2020ApJ...900L...8C, 2021MNRAS.506.4654W}.
In addition,
we adopted the formula in \citet{1971ctf..book.....L} to calculate
the orbital angular momentum carried by the gravitational wave (GW) radiation:
\begin{equation}
\frac{dJ_{\rm GR}}{dt} = -\frac{32G^{7/2}} {5c^5}\, \frac{M_{\rm NS}^2 M_{\rm He}^2 (M_{\rm NS}+M_{\rm He})^{1/2}}
{a^{7/2}},
\end{equation}
where $M_{\rm NS}$ and $M_{\rm He}$ are the mass of NS and the He star companion,
$a$ is the orbital separation of the binary,
$G$ and $c$ denote the gravitational constant and the speed of
light in vacuum, respectively.

For the evolution of He star companions,
we used the `Dutch' stellar wind with a efficiency of $1.0$
to calculate the wind mass-loss rate
\citep[e.g.][]{2009A&A...497..255G,Antoniadis2020A&A,2021ApJ...920L..36J},
and considered the orbital angular momentum loss caused by the stellar wind.
Our nuclear network includes 43 isotopes from $^1\rm H$ to $^{58}\rm Ni$,
involving NeNa and MgAl cycles, Urca processes ($^{23}\rm Na\rightleftharpoons$ $^{23}\rm Ne$,
$^{23}\rm Ne\rightleftharpoons$ $^{23}\rm F$,
$^{25}\rm Mg\rightleftharpoons$ $^{25}\rm Na$ and $^{25}\rm Na\rightleftharpoons$ $^{25}\rm Ne$),
and the electron-capture chains,
i.e. $^{24}\rm Mg$$(e^-,\nu_e)^{24}\rm Na$$(e^-,\nu_e)^{24}\rm Ne$ and
$^{20}\rm Ne$$(e^-,\nu_e)^{20}\rm F$$(e^-,\nu_e)^{20}\rm O$.
We adopted the weak interaction rates provided by \citet{2016ApJ...817..163S}.
The nuclear network does not include silicon burning and nuclear statistical equilibrium.
Following the method described in \citet{2017MNRAS.472.3390S},
we used the HELM \citep{2000ApJS..126..501T} and the PC \citep{2010CoPP...50...82P} equations-of-state,
and ensure that the metal core is treated by the PC equations-of-state.

In our calculations,
we set the mixing-length parameter to be $2.0$,
and applied the Ledoux criterion.
To avoid numerical difficulties for the evolution of the NS binaries,
we used the variant of standard mixing-length theory (i.e. MLT++) in MESA to calculate convective energy transport
\citep[e.g.][]{2015ApJS..220...15P, Antoniadis2020A&A, 2022A&A...668A.106C}.
The overshooting parameter $f_{\rm ov}$ in our models is set to be $0.014$
\citep[e.g.][]{2013ApJ...772..150J,Antoniadis2020A&A}.
Besides,
we considered both semi-convection \citep{1983A&A...126..207L}
and thermohaline mixing \citep{1980A&A....91..175K}.
We did not adopt the predictive convective boundary mixing scheme provided by \citet{2018ApJS..234...34P}.
We used the Type $2$ opacities \citep{1996ApJ...464..943I},
which is applicable to extra carbon and oxygen due to the He-burning.
The physical assumptions for the evolution of He star is similar to the work of
\citet{Antoniadis2020A&A}\footnote{Their MESA inlists
and the related python scripts are publicly available at \href{https://zenodo.org/record/3580243\#.Y9SS0y-KHJN}{https://zenodo.org/record/3580243\#.Y9SS0y-KHJN}.
The MESA inlists in our work are publicly available at
\href{https://zenodo.org/record/7655310\#.Y\_LaNS-KGx9}{https://zenodo.org/record/7655310\#.Y\_LaNS-KGx9.}}.

To explore the formation of SNe Ia through NS+He star systems,
we carried out a large number of complete binary evolution calculations
with different initial He star masses ($M_{\rm He}^{\rm i}=2.5-2.9\rm\,M_\odot$)
and different initial orbital periods ($P_{\rm orb}^{\rm i}=0.07-10$\,d),
in which we used an equal step of $\Delta M_{\rm He}^{\rm i}=0.01\rm\,M_\odot$,
and set $\Delta$log\,$P_{\rm orb}^{\rm i}$ to be $0.5$.\footnote{We calculated a total of $\sim120$ binary models,
in which $70$ models reach explosive oxygen burning, $\sim 25$ models undergo EC-SNe,
and the rest $\sim 25$ models evolve into ONe WDs.}

Following previous studies,
we adopted the point when the temperature starts to increase dramatically
but the density stops changing as the point of the explosive oxygen ignition
\citep[e.g.][]{2006MNRAS.368..187L, 2014MNRAS.438.3358C, 2017MNRAS.472.1593W}.
In our calculations,
the temperature rises sharply
after the explosive oxygen burning or $e$-capture on $\rm^{20}Ne$ occurs,
and thereby resulting in a smaller time-step.
The evolutionary code stops when the time-step reaches the minimum value
(i.e. $10^{-6}$ seconds).
In addition,
our simulations show that the mechanism of triggering oxygen ignition is the convective Urca process (see Section \ref{sec:Urca}).
Thus,
we terminated the code if the mass of ONe core is less than $\sim 1.335\rm\,M_\odot$
because the central density cannot reach the critical value for the Urca process to occur.

\section{Numerical results}

\subsection{A typical example for binary evolution calculations}\label{sec:example}
\begin{figure}
	\centering\includegraphics[width=\columnwidth*5/5]{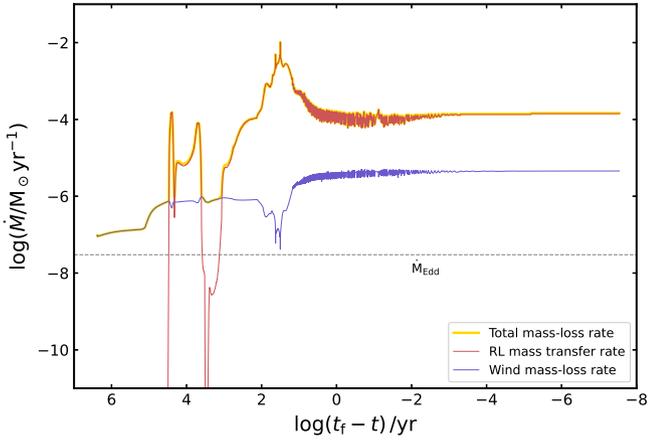}
	\caption{
		Evolutionary track of the mass-loss rate for the $2.65\rm\,M_\odot$ He star companion,
		in which the yellow, red and purple lines
	represent the total mass-loss rate, the mass-transfer rate and the stellar wind mass-loss rate, respectively.
	The value of ($t_{\rm f}-t$) is the remaining time of the He star companion evolution,
	in which the total evolution time $t_{\rm f}\sim2.40$\,Myr.
The dashed line denotes the Eddington accretion rate
($\dot M_{\rm Edd}=3 \times 10^{-8}\,\rm M_\odot \rm yr^{-1}$) assumed in this work.}
	\label{fig:mass-loss}
\end{figure}
\begin{figure}
	\centering\includegraphics[width=\columnwidth*5/5]{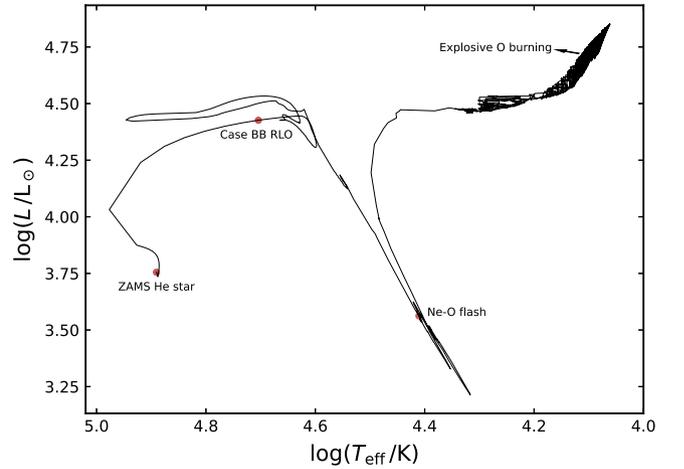}
	\caption{The H-R diagram for the $2.65\,\rm M_\odot$ He star companion from He-ZAMS to explosive oxygen burning,
	where the brown dots indicate the onset of the He star evolution, the Case BB RLOF and the Ne-O flashes.}
	\label{fig:H-R}
\end{figure}
\begin{figure}
	\centering\includegraphics[width=\columnwidth*5/5]{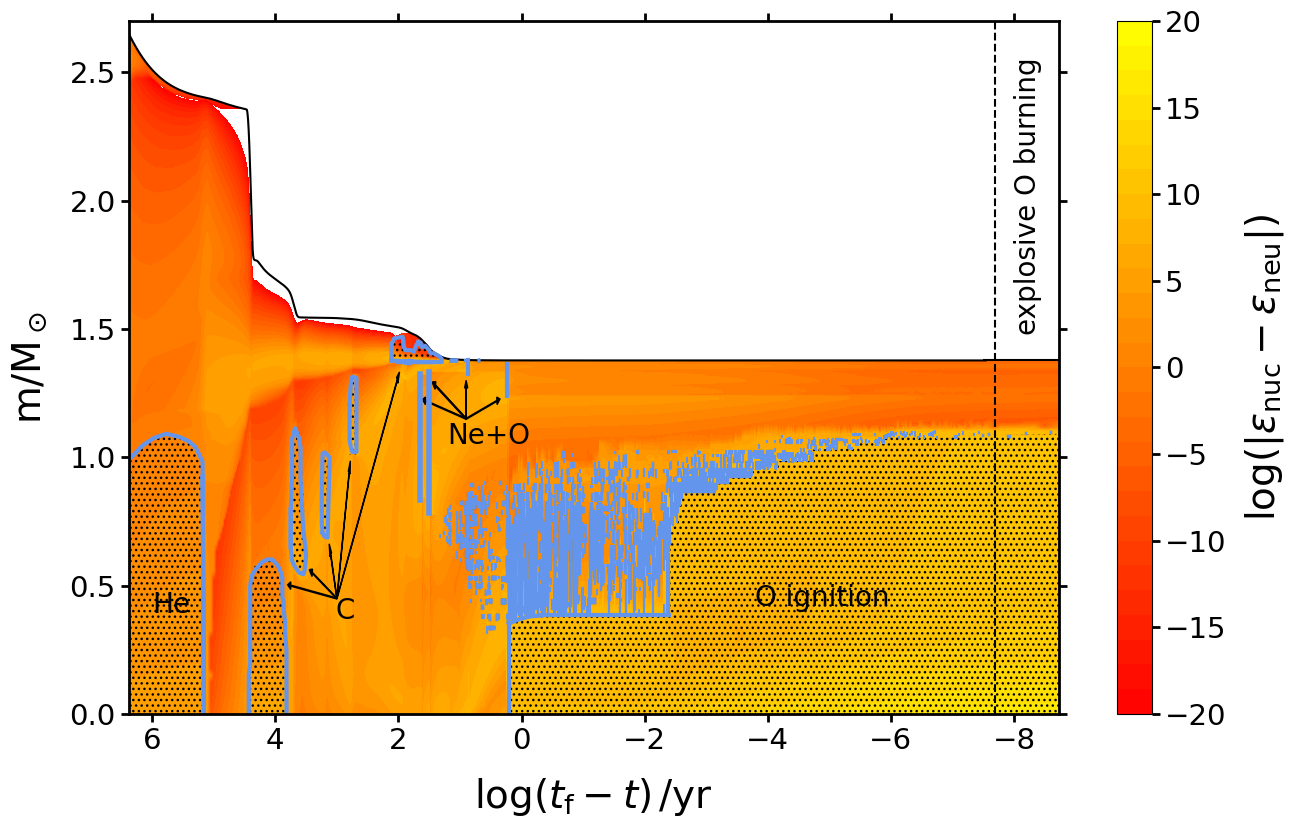}
	\caption{Kippenhahn diagram of the $2.65\rm\,M_\odot$ He star companion from He-ZAMS to explosive oxygen burning,
		including the evolution of interior structure and energy production.
		The hatched regions denote convection
		caused by the He-, C-, and advanced burning phases.
		The blue regions indicate the convection regions.
		The intensity shown in the color-bar represents the nuclear energy-production rate.
	Oxygen ignition is triggered at log$(t_{\rm f}-t)/\rm yr\sim0.2$,
and explosive oxygen burning is triggered at log$(t_{\rm f}-t)/\rm yr\sim-7.7$.}
	\label{fig:kip-yin6-2}
\end{figure}
\begin{figure}
	\centering\includegraphics[width=\columnwidth*5/5]{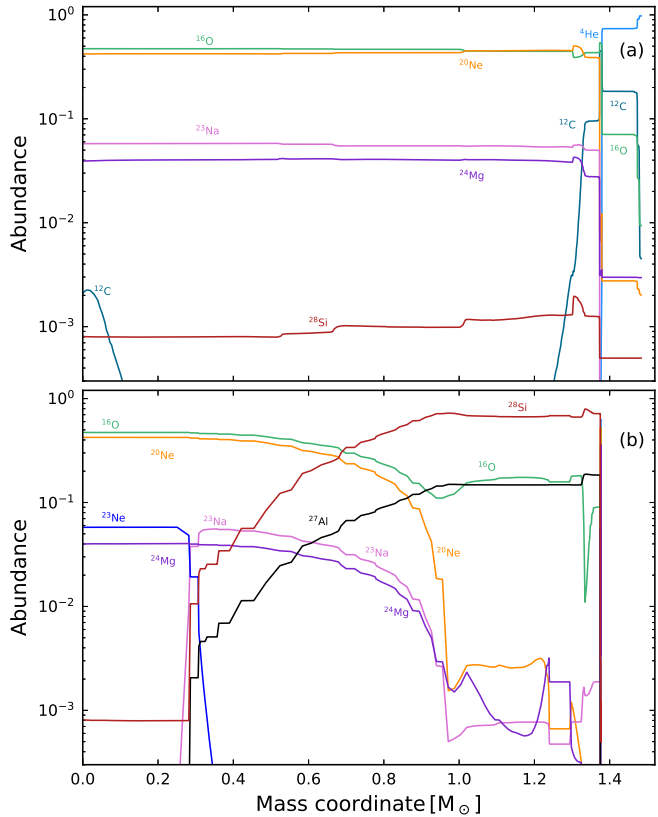}
	\caption{Panel (a): the chemical structure of the donor at the end of carbon burning stage.
		At this moment,
		the donor mass is $\sim 1.46\,\rm M_\odot$,
		including a $\sim 1.378\,\rm M_\odot$ ONe core and a $\sim 0.082\,\rm M_\odot$ He-layer.
		The ONe core is mainly composed of oxygen, neon, sodium and magnesium with masses of
		$M_{\rm c}(^{16}\rm O)\approx0.65\,\rm M_\odot$, $M_{\rm c}(^{20}\rm Ne)\approx0.58\,\rm M_\odot$,
		$M_{\rm c}(^{23}\rm Na)\approx0.08\,\rm M_\odot$ and $M_{\rm c}(^{24}\rm Mg)\approx0.06\,\rm M_\odot$,
		as well as the residual carbon with mass of $M_{\rm c}(^{12}\rm C)\approx0.0002\,\rm M_\odot$.
		Panel (b): the chemical structure of the donor after Ne-O flashes,
		in which the metal core consists of an ONe core and a thick Si-rich mantle.
}
	\label{fig:ele-one}
\end{figure}
\begin{figure}
	\centering\includegraphics[width=\columnwidth*5/5]{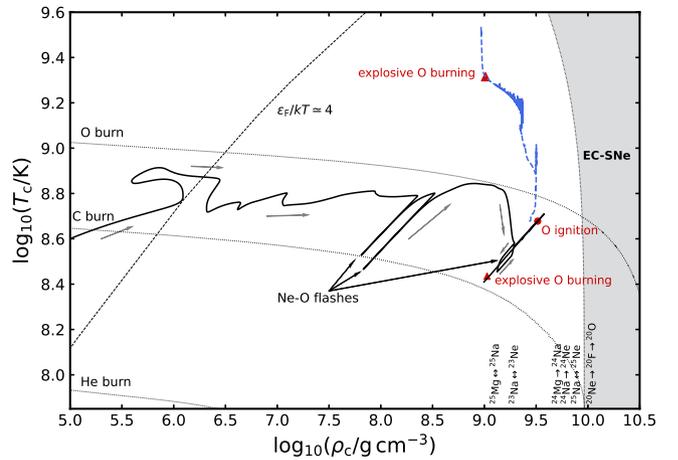}
	\caption{Evolutionary track of the central temperature and the central
		density for the $2.65\,\rm M_\odot$ He star companion
		(the black line),
		where the tracks follow the direction of the gray arrows.
		The red circle and the red triangle denote
		the moment of the oxygen ignition and the explosive oxygen burning occur, respectively.
		Grey region indicates the region that can evolve to EC-SNe.
		The oxygen ignition is off-center in this model.
		Thus, we plot the evolutionary track of the temperature and density at oxygen ignition site to prove that the explosive oxygen burning occurs (the blue dashed line).
The dotted lines represent helium, carbon and oxygen burning ignition curves,
and the dashed line denotes the separation of degenerate and non-degenerate regions
($\epsilon_{\rm F}/k T \simeq 4$).
}
	\label{fig:pc-tc1}
\end{figure}
\begin{figure}
	\centering\includegraphics[width=\columnwidth*5/5]{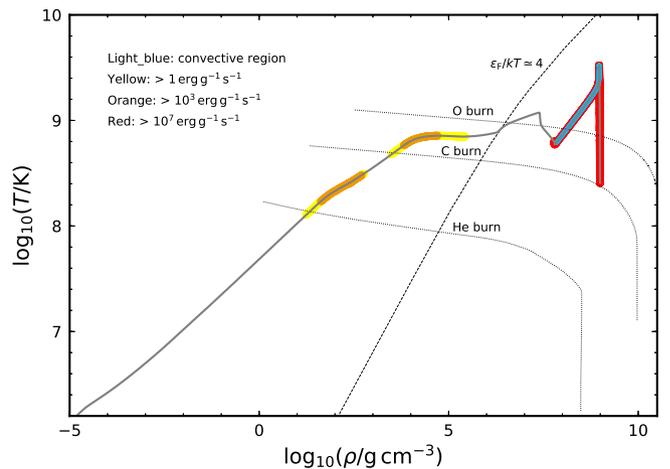}
	\caption{Profile of $\rho$ and $T$ of the mass donor from our final model at $t \sim 2.396407$\,Myr,
		in which the explosive oxygen burning is triggered at $M_{\rm c}\approx 0.0008\,\rm M_\odot$.
	The yellow, brown and red colors represent different energy production rates,
	and the light blue color denotes the convective region.}
	\label{fig:p-t}
\end{figure}
\begin{figure}
	\centering\includegraphics[width=\columnwidth*5/5]{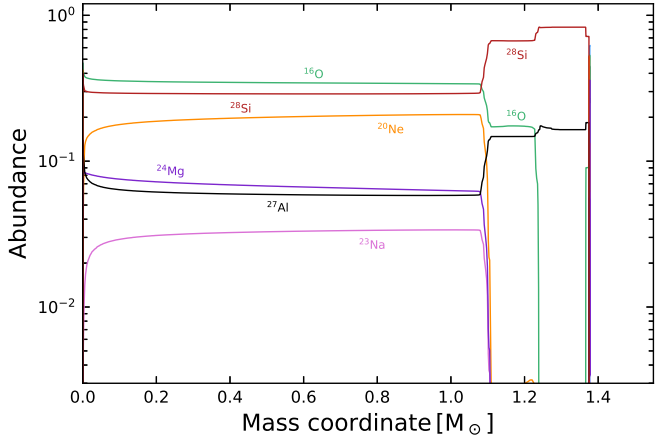}
	\caption{The chemical structure of the mass donor from the last model at $t \sim 2.396407$\,Myr,
		in which the explosive oxygen burning is initiated.
	}
	\label{fig:elef}
\end{figure}

We present a typical example of the evolution of a NS+He star system
that can form an SN Ia (see Figs\,\ref{fig:mass-loss}$-$\ref{fig:elef}),
in which the initial parameters for this binary are
$M_{\rm NS}^{\rm i} = 1.35\rm\,M_\odot$,
$P_{\rm orb}^{\rm i}=1.0$\,d and $M_{\rm He}^{\rm i} = 2.65\rm\,M_\odot$.
Fig.\,\ref{fig:mass-loss} shows the mass-loss rate of the He star companion
as a function of remaining time.
The total evolutionary time from He-ZAMS to the explosive oxygen burning is $\rm \sim2.40\,Myr$.
Before the companion fills its Roche lobe,
the mass loss for the companion is caused by the stellar wind.
When the binary age $t\sim2.36$\,Myr,
Roche-lobe overflow (RLOF) is initiated due to the He star expansion after core He exhaustion
(i.e. Case BB RLOF),
triggering the stable mass transfer from He star to NS
\citep{2015MNRAS.451.2123T}.
After the onset of Case BB RLOF,
the mass transfer dominates the mass loss of the donor.
During the mass-transfer stage,
the binary appears to be a strong X-ray source with luminosity $\rm \sim 10^{38}\,erg\,s^{-1}$.
Fig.\,\ref{fig:H-R} represents the complete evolutionary track of the He star companion in the H-R diagram.
We marked the moment when the He star evolution, the Case BB RLOF, the Ne-O flashes and explosive oxygen burning begin.


Fig.\,\ref{fig:kip-yin6-2} shows the Kippenhahn diagram of the He star companion
from He-ZAMS to explosive oxygen burning,
including the evolution of interior structure and energy production.
After core He depletion at $t\sim2.24$\,Myr,
He-shell burning take places,
resulting in the expansion of the envelope.
Following the core He burning phase,
a CO core is formed and gradually contracts
until the ignition of the central carbon at $t\sim2.37$\,Myr.
After the exhaustion of carbon in the center,
carbon burning in shell appears and gradually propagates outward.

Fig.\,\ref{fig:ele-one}a illustrates the chemical structure of the mass donor
at the end of carbon burning stage ($t\sim2.396355$\,Myr).
At this moment,
the donor decreases its mass to $\sim 1.46\,\rm M_\odot$.
Meanwhile,
the donor is composed of a $\sim 1.378\,\rm M_\odot$ ONe core and a $\sim 0.082\,\rm M_\odot$ He-layer,
in which the mass of the ONe core exceeds the critical value for neon ignition
\citep[$\sim 1.37\,\rm M_\odot$, e.g.][]{Nomoto1984, 2013ApJ...771...28T}.
The ONe core is mainly composed of oxygen, neon, sodium and magnesium with
masses of $M_{\rm c}(^{16}\rm O)\approx0.65\,\rm M_\odot$, $M_{\rm c}(^{20}\rm Ne)\approx0.58\,\rm M_\odot$,
$M_{\rm c}(^{23}\rm Na)\approx0.08\,\rm M_\odot$ and $M_{\rm c}(^{24}\rm Mg)\approx0.06\,\rm M_\odot$.
In addition,
the amount of residual carbon in degenerate core after carbon burning is $M_{\rm c}(^{12}\rm C)\approx0.0002\,\rm M_\odot$,
which is significantly lower than $M_{\rm c}(^{12}\rm C)\gtrsim0.003\,\rm M_\odot$
proposed by \citet{2022A&A...668A.106C},
see discussion in Section \ref{sec:comp}.

Neon is ignited off-center ($t \sim 2.396364$\,Myr)
at mass coordinates $M_{\rm c}$ of $\sim 0.87\,\rm M_\odot$
(see Fig.\,\ref{fig:kip-yin6-2}).
This is because neutrino emission processes
remove the energy generated by gravitational contraction,
resulting in a temperature inversion in the metal core
\citep[e.g.][]{Nomoto1984, 2013ApJ...772..150J}.
Besides, the Ne-shell burning is unstable to a flash
because of the degenerate shell.
Subsequently,
oxygen is ignited in the shell because the Ne flash increases the temperature.
Fig.\,\ref{fig:pc-tc1} represents the evolutionary track of central temperature ($T_{\rm c}$) and
central density ($\rho_{\rm c}$) for the donor (the black line).
During the Ne-O flashes,
neon and oxygen burn into silicon and other heavy elements.
Meanwhile,
the shell-burning layer expands,
and then the pressure on the central region decreases.
As a result, $T_{\rm c}$ and $\rho_{\rm c}$ decrease
because of the almost adiabatic expansion of the metal core.
After the Ne-O flash quenches,
the shell contracts and provides a higher pressure to the core,
resulting in a higher $T_{\rm c}$ and $\rho_{\rm c}$
\citep[][]{2013ApJ...772..150J}.
As the central density gradually increases,
the Urca reactions $^{25}\rm Mg\rightleftharpoons$ $^{25}\rm Na$ and
$^{23}\rm Na\rightleftharpoons$ $^{23}\rm Ne$
are triggered at $\rm log_{10}(\rho_c/g\,cm^{-3})\approx9.1$ and $\rm log_{10}(\rho_c/g\,cm^{-3})\approx9.2$,
accelerating the core contraction toward higher $\rho_{\rm c}$.
As the metal core continues to shrink,
the $e$-captures on $^{24}\rm Mg$ takes place at $\rm log_{10}(\rho_c/g\,cm^{-3})\approx9.6$.
Fig.\,\ref{fig:ele-one}b shows the chemical structure of the donor after Ne-O flashes.
The residual carbon at the central core and the outer core
is gradually consumed during the Ne-O flashes.

Following a few Ne-O shell flashes,
the oxygen ignition (log$(t_{\rm f}-t)\,/\rm yr\approx-0.2$) is triggered near the center
at $M_{\rm c}$ of $\sim 0.0008\,\rm M_\odot$,
rather than the $e$-captures on $^{20}\rm Ne$.
The blue dashed line in Fig.\,\ref{fig:pc-tc1} represents
the evolutionary track of the temperature and density at oxygen ignition site.
At the moment of log$(t_{\rm f}-t)\,/\rm yr\approx-7.7$,
the temperature at the oxygen ignition site starts to increase sharply,
indicating that the explosive oxygen burning is triggered (see Fig.\,\ref{fig:kip-yin6-2}).
Fig.\,\ref{fig:p-t} illustrates the $\rho-T$ profile of the mass donor
in our final model ($t \sim 2.396407$\,Myr),
where the maximum temperature reaches $3.31\times10^9$\,K,
and the metal core develops a large convective region from
$M_{\rm c}\sim0.0008\,\rm M_\odot$ to $1.07\,\rm M_\odot$
(see Fig.\,\ref{fig:kip-yin6-2}).
Fig.\,\ref{fig:elef} shows the chemical structure of the mass donor in the final model.
At this moment, the metal core consists of an O-Ne-Si core and a thick Si-rich mantle.
Meanwhile,
the donor star has $\sim0.4\,\rm M_\odot$ of $\rm^{16}O$ and $\sim0.22\,\rm M_\odot$ of $\rm^{20}Ne$,
as well as a binding energy of $-5.25\times10^{50}$\,erg.
If we assume that the thermonuclear runaway produces relatively low mass nickel
\citep[$0.2\,\rm M_\odot$,][]{Jones2019A&A},
then the nuclear energy generated is $\sim6.52\times10^{50}$\,erg.
Accordingly, the donor could be completely disrupted,
and the kinetic energies for the ejecta is about $1.27\times10^{50}$\,erg.

\subsection{Mechanism of triggering oxygen ignition}\label{sec:Urca}

\begin{figure}
	\centering\includegraphics[width=\columnwidth*5/5]{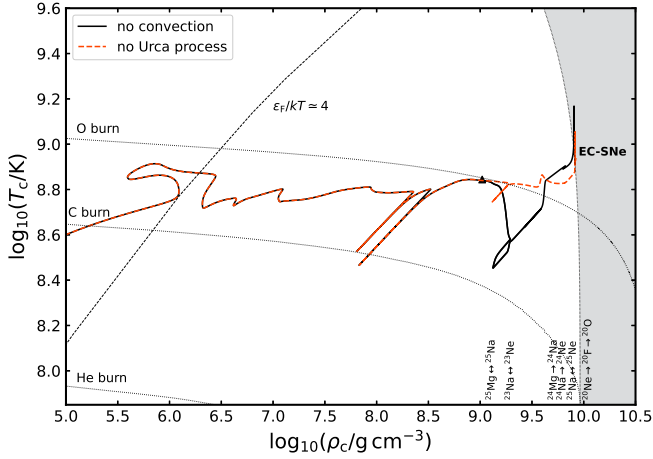}
	\caption{Evolutionary tracks of two variations of the typical example model in the log$T_{\rm c}$ $-$ log$\rho_{\rm c}$ diagram.
		One model does not undergo the Urca process cooling (red dashed line).
		In another model (black line), we artificially removed the convective regions in the degenerate core
		when the central density reaches $\sim 10^{9}\rm \,g\,cm^{-3}$ (black triangle).
	}
	\label{fig:pc-tc1-nm}
\end{figure}
Following the method described in \citet{Antoniadis2020A&A} and \citet{2022A&A...668A.106C},
we set all carbon-participating nuclear reactions to be zero
after the central density became larger than $\sim 10^{9}\rm \,g\,cm^{-3}$.
We found that oxygen can still be ignited,
indicating that the oxygen ignition is not caused by the residual carbon burning.
We checked nuclear reactions other than the residual carbon burning,
and found that the mechanism of triggering oxygen ignition is convective Urca process,
which can result in substantial net heating
\citep[e.g.][]{2005MNRAS.356..131L,2008NewAR..52..381P,2017ApJ...851..105S}.
During the Urca-process cooling,
convection occurs in the region that has experienced Urca process.
If the convective region grows to a threshold density spanning one or more Urca pairs,
convection can transport material undergoing $e$-capture in the higher density regions to the lower density regions,
thereby triggering the beta-decay reactions
\citep[i.e. $^{25}\rm Na\rightarrow$ $^{25}\rm Mg$ and $^{23}\rm Ne\rightarrow$ $^{23}\rm Na$;][]{2017MNRAS.472.3390S}.

Fig.\,\ref{fig:pc-tc1-nm} shows the evolutionary tracks of central temperature ($T_{\rm c}$) and
central density ($\rho_{\rm c}$) for the He star companions in the two variations of the typical example model,
where neither model experienced the convective Urca process.
We can see that both models reach the EC-SN stage instead of the ignition of oxygen.
This means that in the absence of the convective Urca process,
the He star companion would produce a core-collapse
if the $e$-capture rate exceeds the nuclear burning rate.
Meanwhile,
our simulations demonstrate that
the convective Urca process plays an important role during the evolution of the degenerate ONe core,
although this heating mechanism still needs further investigation
\citep[e.g.][]{2005MNRAS.356..131L,2008NewAR..52..381P,2017MNRAS.472.3390S,2017ApJ...851..105S}.

\subsection{Parameter space for SNe Ia}
We calculated a series of NS+He star binaries with different
$M_{\rm He}^{\rm i}$ and $P_{\rm orb}^{\rm i}$,
and thereby obtained an initial parameter space for the formation of SNe Ia.
Table\,\ref{table:1} lists the initial input parameters and
main evolutionary properties about the NS+He star systems
at the upper and lower boundaries of the parameter space for SNe Ia.
The simulations show that the metal core masses
for producing SNe Ia in NS+He star binaries range from $\sim 1.335\rm\,M_\odot$ to $\sim 1.385\rm\,M_\odot$.
If the metal core mass $\lesssim 1.335\rm\,M_\odot$,
then the convective Urca process will not be triggered
because the central density cannot reach the critical value for the Urca process.
If the core mass $\gtrsim 1.385\rm\,M_\odot$,
then $e$-capture on $\rm^{20}Ne$ will be triggered.
In addition, the $\rho_{\rm c}$ for oxygen ignition
is in the range of $\sim1.3-6.5\times10^9\rm\,g\,cm^{-3}$.
\begin{table*}
	\centering
	
	\caption{
		Information about the NS+He star systems at the upper and lower boundaries of the parameter space for the SNe Ia.
		$M_{\rm He}^{\rm i}$ and log$P_{\rm orb}^{\rm i}$ are the initial He star mass in solar masses
		and the initial orbital period in days;
		$M_{\rm core, f}$ is the final metal core mass of the donor;
		$M_{\rm c}(^{12}\rm C)$ is the total amount of carbon in the degenerate core after carbon burning phase;
		$\Delta M_{\rm NS}$ and $\Delta t_{\rm X}$ are the accreted mass of NS and
		the duration of the X-ray phase;
		$P_{\rm spin}^{\rm min}$ is the minimum spin period of NS;
		$P_{\rm orb}^{\rm f}$ and $V_{\rm orb}^{\rm f}$ are the final orbital period and
		the final orbital velocity of NS prior to SN;
		$t_{\rm f}$ is the evolution time of He star companions from He-ZAMS to explosive oxygen burning;
		the last column is the final fate of the He star companion.
	}
	\label{table:1}
	\begin{tabular}{ c  c c c cccc c c c cc c }
		\toprule
		\hline 
		Set		&&$M_{\rm He}^{\rm i}$  &log$P_{\rm orb}^{\rm i}$  &$M_{\rm core, f}$&$M_{\rm c}(^{12}\rm C)$ & $\Delta M_{\rm NS}$ &$\Delta t_{\rm X}$
		&$P_{\rm spin}^{\rm min}$ &$P_{\rm orb}^{\rm f}$ &$V_{\rm orb}^{\rm f}$ &$t_{\rm f}$	&Final fate\\
		&&($\rm M_\odot$)&(d)& ($\rm M_\odot$)&($\rm M_\odot$) & $(\rm M_\odot)$  &(yr) & (ms) &(d)& ($\rm km\,s^{-1}$)&(Myr) \\
		\hline 
		1	&&		$2.84$	&$-1.15$&$1.381$ &1.6e-4	&2.0e-3	&9.0e4	&$36.0$		&$0.075$ 	&$356$	&$2.14$	&EC-SN\\
		2	&&		$2.74$ 	&$-1.15$&$1.338$ &7.5e-4 	&2.3e-3	&1.0e5	&$32.4$		&$0.080$	&$342$	&$2.26$	&SN Ia\\
		\hline 
		3	&&		$2.79$ 	&$-1.00$&$1.389$ &1.0e-4	&1.7e-3 &7.7e4	 &$40.6$	&$0.113$	&$312$	&$2.20$	&EC-SN\\
		4	&&		$2.67$ 	&$-1.00$&$1.337$ &8.2e-4	&2.0e-3 &8.9e4	 &$36.0$	&$0.121$	&$298$	&$2.37$	&SN Ia\\
		\hline 
		5	&&		$2.70$	&$-0.50$&$1.396$ &1.0e-4	&8.7e-4	&4.1e4	 &$67.1$	&$0.378$	&$209$	&$2.32$	&EC-SN\\
		6	&&		$2.57$	&$-0.50$&$1.335$ &1.0e-3	&4.1e-3	&5.1e4	 &$56.3$	&$0.406$	&$199$	&$2.52$	&SN Ia\\
		\hline
		7	&&		$2.67$	&$0$	&$1.385$ &1.6e-4	&8.2e-4	&3.0e4	 &$70.2$	&$1.226$	&$140$	&$2.37$	&EC-SN\\
		8	&&		$2.55$	&$0$	&$1.336$ &1.0e-3	&7.0e-4	&3.4e4	 &$79.0$	&$1.301$	&$135$	&$2.55$	&SN Ia\\
		\hline
		9	&&		$2.65$	&$0.5$	&$1.383$ &1.7e-4	&6.7e-4	&2.5e4	 &$81.6$	&$3.915$	&$95$	&$2.40$	&EC-SN\\
		10	&&		$2.54$	&$0.5$	&$1.336$ &1.0e-3	&7.7e-4	&2.9e4	 &$73.6$	&$4.142$	&$92$	&$2.60$	&SN Ia\\	
		\hline
		11	&&		$2.66$	&$1$	&$1.384$ &1.5e-4	&4.8e-4	&2.0e4	 &$104.8$	&$12.39$	&$65$	&$2.38$	&EC-SN\\
		12	&&		$2.54$	&$1$	&$1.338$ &8.6e-4	&6.7e-4	&2.7e4	 &$81.6$	&$13.04$	&$63$	&$2.60$	&SN Ia\\	
		\hline
	\end{tabular}
\end{table*} 

\begin{figure}
	\centering\includegraphics[width=\columnwidth*5/5]{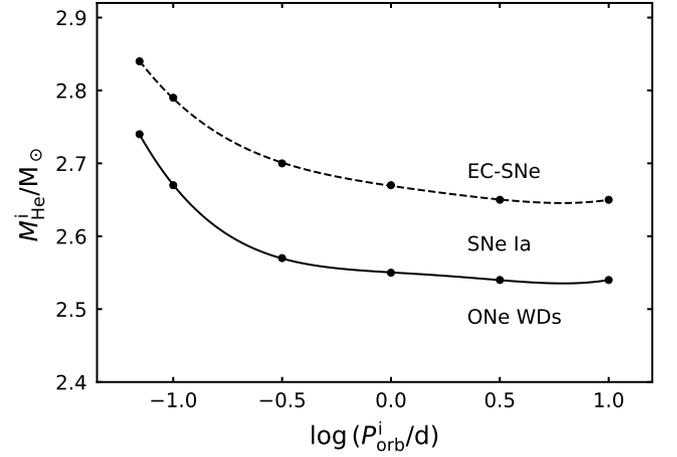}
	\caption{Initial parameter space for producing SNe Ia in the
		$\rm log\,$$P^{\rm i}_{\rm orb}-M^{\rm i}_{\rm He}$ plane.
		The dashed line indicates the boundary between SNe Ia and EC-SNe,
		and the solid line represents the boundary between SNe Ia and ONe WDs.
	}
	\label{fig:Ia}
\end{figure}
Fig.\,\ref{fig:Ia} shows the initial contour for producing SNe Ia
in the $\rm log\,$$P^{\rm i}_{\rm orb}-M^{\rm i}_{\rm He}$ plane.
He star companions can explode as SNe Ia if the initial parameters of
NS+He star systems are located in this parameter space.
The zero-age He MS star will fill its Roche lobe if $P^{\rm i}_{\rm orb}\lesssim0.07$\,d,
and the He star companions would have similar final outcomes if $P^{\rm i}_{\rm orb}\gtrsim10$\,d.
The black dashed line represents the boundary between SNe Ia and EC-SNe,
and the black solid line represents the boundary between SNe Ia and ONe WDs.
We used polynomial function to fit the relationship between $M^{\rm i}_{\rm He}$
and $\rm log\,P^{\rm i}_{\rm orb}$ at the upper and lower boundaries,
which can be used for population synthesis studies.
For upper boundary:
\begin{equation}
	\begin{split}
	M^{\rm i}_{\rm He} = &0.032\rm (log\,P^{\rm i}_{\rm orb})^4-0.025\rm (log\,P^{\rm i}_{\rm orb})^3\\
     &+0.018\rm (log\,P^{\rm i}_{\rm orb})^2-0.044\rm (log\,P^{\rm i}_{\rm orb})+2.669,
   \end{split}
\end{equation}
for lower boundary:
\begin{equation}
	\begin{split}
	M^{\rm i}_{\rm He} = &0.05\rm (log\,P^{\rm i}_{\rm orb})^4-0.047\rm (log\,P^{\rm i}_{\rm orb})^3\\
	&+0.004\rm (log\,P^{\rm i}_{\rm orb})^2-0.018\rm (log\,P^{\rm i}_{\rm orb})+2.55.
    \end{split}
\end{equation}

\citet{2022A&A...668A.106C} evolved a series of single He stars
with different metallicities from $10^{-4}$ to $0.02$.
Their simulations indicate that lower metallicities result in more massive metal cores.
This may cause the initial parameter space for SNe Ia to move down in this work
if a low value of metallicities is adopted.
Meanwhile, lower metallicities can also lead to smaller initial radius of He stars,
resulting in that the initial parameter space moves to left slightly
\citep[e.g.][]{2010A&A...515A..88W, 2022A&A...668A.106C}.

\begin{figure}
	\centering\includegraphics[width=\columnwidth*5/5]{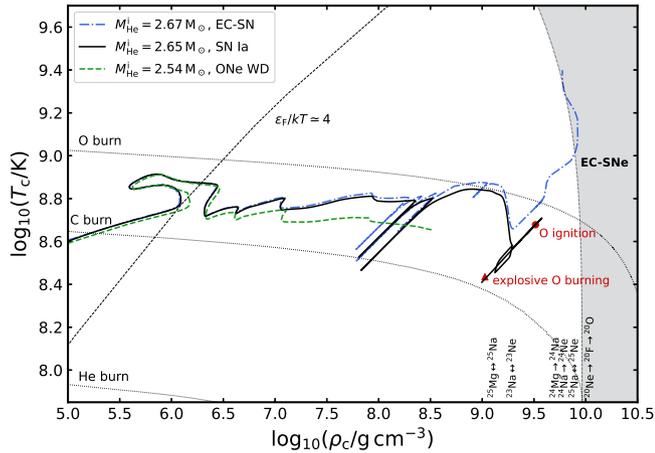}
	\caption{Evolutionary tracks of the central temperature and the central density
		for He star companions with different initial masses,
		in which the initial orbital period is $1.0$\,d.
			The red circle and the red triangle denote
			the moment of the oxygen ignition and the explosive oxygen burning occur, respectively.}
	
	\label{fig:pc-tc}
\end{figure}

Fig.\,\ref{fig:pc-tc} shows the representative examples of three NS+He star systems
with different outcomes
in the $\rho_{\rm c}-T_{\rm c}$ plane,
i.e. ONe WD, SN Ia and EC-SN.
For NS+He star binary with $M^{\rm i}_{\rm He}=2.54\,\rm M_\odot$,
the final metal core mass is $\sim 1.330\,\rm M_\odot$.
The central density does not reach the threshold value for the occurrence of the Urca reaction,
thus the convective Urca process will not be triggered,
and this He star companion evolves into an ONe WD eventually.
For NS+He star binary with $M^{\rm i}_{\rm He}=2.65\,\rm M_\odot$,
the final metal core mass is $\sim 1.378\,\rm M_\odot$,
explosive oxygen burning can be triggered due to
convective Urca process.
For NS+He star binary with $M^{\rm i}_{\rm He}=2.67\,\rm M_\odot$,
the final metal core mass is $\sim 1.385\,\rm M_\odot$.
The central density increases relatively quickly,
resulting in that the $e$-capture on $^{20}\rm Ne$ occurs before the significant convective Urca process,
and therefore the formation of an EC-SN.

\begin{figure*}
	\centering\includegraphics[width=\columnwidth*9/5]{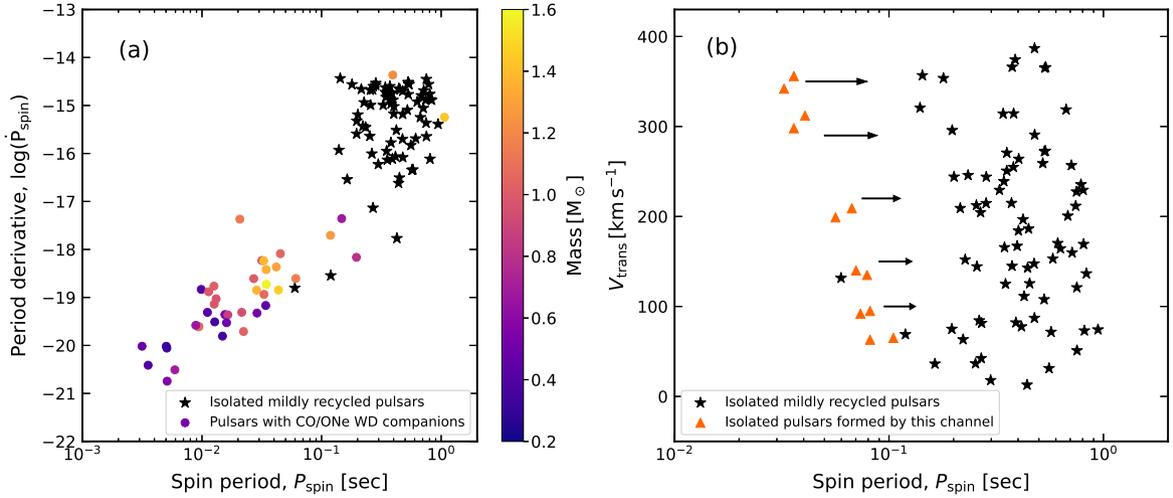}
	\caption{Comparison between our simulation results and observations.
		Panel (a): $P_{\rm spin}$ vs. $\dot P_{\rm spin}$ for Galactic disk pulsars with
		CO/ONe WD companions (circles), and for isolated mildly recycled pulsars (black stars) with
		$P_{\rm spin}$ from $0.01$\,s to $1$\,s and $\dot P_{\rm spin}<4\times10^{-15}\rm\,s\,s^{-1}$.
		The intensity indicates the median WD companion mass corresponding to
		an orbital inclination angle of $60^{\circ}$.
		Panel (b): $P_{\rm spin}$ vs. $V_{\rm trans}$ for
		the isolated mildly recycled pulsars (black stars) in Galactic disk.
		Orange triangles represent the isolated pulsars formed through the NS+He star channel,
		in which we take the final orbital velocities of pulsars as their transverse velocities,
		and the data were taken from Table\,\ref{table:1}.
		The observed data of pulsars were taken from the ATNF Pulsar Catalogue, \href{http://www.atnf.csiro.au/research/pulsar/psrcat}{http://www.atnf.csiro.au/research/pulsar/psrcat}
		\citep[version 1.68, October 2022;][]{2005AJ....129.1993M}.
	}
	\label{fig:imsp}
\end{figure*}
\section{Isolated mildly recycled pulsars}
\subsection{Spin period and final orbital velocity for pulsars}\label{sec:spin}
It is generally believed that
NSs in low-mass X-ray binaries can spin up by accreting the material
and angular momentum from their companions
\citep[e.g.][]{1982Natur.300..728A, 1991PhR...203....1B}.
We adopted the equation provided by \citet{2012MNRAS.425.1601T}
to calculate the minimum spin period ($P_{\rm spin}^{\rm min}$) of the recycled NSs,
which is given by
\begin{equation}
	P_{\rm spin}^{\rm min} \approx 0.34\times(\Delta M_{\rm NS}/\rm M_\odot)^{-3/4}\,ms,
\end{equation}
where $\Delta M_{\rm NS}$ is the accreted mass of NS.

We found that $\Delta M_{\rm NS}$
calculated in this work is in the range of
$0.5-2.3\times10^{-3}\rm\,M_\odot$ (see Table\,\ref{table:1}).
This may lead to the formation of isolated mildly recycled pulsars
with $P_{\rm spin}^{\rm min}\sim30-110$\,ms
after the He stars undergo the SN Ia explosions.
In addition, we note that shorter $P_{\rm orb}^{\rm i}$ (or lower $M_{\rm He}^{\rm i}$)
will result in longer
duration of the accretion phase,
thereby higher accreted masses of NSs
\citep[see also][]{2015MNRAS.451.2123T}.
We also calculated the final orbital velocity of NS prior to SN ($V_{\rm orb}^{\rm f}$),
and found that $V_{\rm orb}^{\rm f}$
calculated in this work ranges from $\sim 60\,\rm km\,s^{-1}$ to $360\,\rm km\,s^{-1}$
(see Table\,\ref{table:1}).
Meanwhile,
a low value of $P_{\rm orb}^{\rm i}$ will result in a higher value of $V_{\rm orb}^{\rm f}$
because of the short final orbital period.
Although the SN ejecta will interact with pulsars,
it has little effect on the space velocity
due to the small radius of the pulsars.
Thus, the orbital velocity of pulsars prior to the SN
can be roughly regarded as the space velocity of pulsars
after He star companions explode as SNe Ia.
\subsection{Comparison with observations}
Our simulations indicate that SNe Ia may be triggered
if the metal core mass is around $M_{\rm Ch}$ in the NS +He star binaries,
resulting in the formation of isolated mildly recycled pulsars.
Meanwhile, in classical binary evolution theory,
the mass transfer in intermediate-mass X-ray binaries
with a intermediate-mass donor star is highly super-Eddington,
thus most of the transferred material may be ejected by radiation pressure,
leaving behind mildly recycled pulsars with massive WD companions eventually
\citep[][]{2000ApJ...530L..93T,2012MNRAS.425.1601T,2002ApJ...565.1107P}.
Accordingly, we expect that such isolated pulsars produced by this channel
may mainly locate in the region of pulsars with massive WD companions
in the $P_{\rm spin}-\dot P_{\rm spin}$ diagram.

In order to compare our results with observations,
we collected all currently known pulsars with CO/ONe WD companions,
and 76 isolated mildly recycled pulsars
that may be produced by this channel,
in which all pulsars are in Galactic disk and
the pulsar data were taken from the ATNF Pulsar Catalogue
\citep[][]{2005AJ....129.1993M}.
Since the upper limits of $P_{\rm spin}$ and $\dot P_{\rm spin}$
for pulsars with WD companions are $\rm \sim 1\,s$ and $\rm 4\times10^{-15}\,s\,s^{-1}$,
the isolated mildly recycled pulsars we collected have $P_{\rm spin}$ between $0.01-1$\,s
and $\dot P_{\rm spin}<4\times10^{-15}\rm\,s\,s^{-1}$.
In addition,
transverse velocity ($V_{\rm trans}$) is
an important parameter for isolated pulsars,
which can be used to reveal the properties of
SN progenitors and explosions, e.g. NS kick velocity
\citep[e.g.][]{2017A&A...608A..57V,2021ApJ...920L..37W,2022ApJ...931..123L}.
We roughly use the maximum value of $V_{\rm orb}^{\rm f}$ calculated in this work
to limit the transverse velocity of the pulsar samples, that is,
these isolated pulsar samples have $V_{\rm trans}$ less than $\rm 400\,km\,s^{-1}$.

Fig.\,\ref{fig:imsp} represents the comparison between simulation results and observations.
In panel (a), we show the $P_{\rm spin}-\dot P_{\rm spin}$ diagram
for pulsars with CO/ONe WD companions (circles),
and for 76 isolated mildly recycled pulsars in Galactic disk (black stars).
We note that such isolated pulsars mainly locate in
the region of pulsars with massive WD companions.
Moreover,
we can see that the observed pulsars with massive WD companions are
preferentially accompanied by higher values of $P_{\rm spin}$,
which is consistent with our results (see Section \ref{sec:spin}).
Panel (b) represents the $P_{\rm spin}-V_{\rm trans}$ diagram
for the isolated mildly recycled pulsars in the observations
and our simulations (orange triangles).
In this work,
we take the calculated final orbital velocities of simulated pulsars
as their transverse velocities (see Table\,\ref{table:1}).
We can see that our simulations can reproduce the transverse velocities
for the isolated mildly recycled pulsars.
Meanwhile,
since the predicted minimum spin periods for pulsars calculated in this work
range from $\sim 30$\,ms to $110$\,ms,
this work has the potential ability to explain the
observed pulsars with relatively longer spin periods,
i.e. the isolated mildly recycled pulsars to the right of the orange triangles
in the $P_{\rm spin}-V_{\rm trans}$ diagram.

\citet{1993ApJ...412L..37C} proposed a high-mass X-ray binary (HMXB) channel
to explain such isolated pulsars, that is,
after the companion undergoes a supernova explosion and forms a newly-born NS,
the binary may be disrupted, leaving behind an isolated mildly recycled pulsar,
called `disrupted recycled pulsars'
\citep[][]{2004MNRAS.347L..21L}.
To produce a relatively high transverse velocity $\gtrsim300\,\rm km\,s^{-1}$,
the pre-SN systems in HMXB channel should possess a short orbital period of
${\rm log}(P_{\rm orb}/\rm d)\lesssim-1$.
However,
the disrupted probability of HMXBs with
${\rm log}(P_{\rm orb}/\rm d)\lesssim-1$ is zero
if the NS kick velocity is less than $220\rm \,km\,s^{-1}$
\citep[see Fig.\,18 in][]{2017ApJ...846..170T}.
Thus, it seems that the HMXB channel may be difficult to explain the formation of
isolated mildly recycled pulsars
with $V_{\rm trans}\gtrsim300\,\rm km\,s^{-1}$.
However,
the origin of these sources can be easily interpreted by our NS+He star channel. 
Accordingly, the isolated mildly recycled pulsars with
relatively high transverse velocity
may preferentially originate from the NS+He star binaries
that can undergo SNe Ia.
In order to verify the NS+He star channel for producing SNe Ia,
more numerical simulations and observations
for SNe Ia or isolated mildly recycled pulsars are needed.

\section{Discussion}
\subsection{Uncertainties}
In this work, we assume that the overshooting parameter $f_{\rm ov}$ to be $0.014$.
However, different values of $f_{\rm ov}$ will affect the final evolutionary products.
\citet{2022A&A...668A.106C} calculated a large number of single He stars
with different values of $f_{\rm ov}$,
and found that higher values of $f_{\rm ov}$
will result in more massive metal cores.
In addition,
the mass loss of He stars is caused by stellar wind
before the Case BB RLOF occurs.
This means that the stellar wind efficiency can also affect the final core mass,
although this effect is not significant \citep[see Fig.\,13 in][]{2022A&A...668A.106C}. 
Accordingly,
the upper and lower boundaries of the parameter space for SNe Ia will be
shifted up or down slightly if different input parameters are adopted.

Some recent works have studied ONe explosions through multidimensional simulations
\citep[][]{2015A&A...580A.118M,Jones2016}.
\citet{2015A&A...580A.118M} explored sub-$M_{\rm Ch}$ ONe WD detonations
at relatively low ignition densities ($\rm \sim 1.0-2.0\times10^8\,g\,cm^{-3}$),
and suggested that the ONe explosions are similar to normal SNe Ia.
However, by simulating the oxygen deflagration at relatively high ignition densities
$\rm log_{10}(\rho_{\rm c}/g\,cm^{-3})\geq9.9$,
\citet{Jones2016} found that about $0.1-1.0\rm\,M_\odot$ of material will be ejected,
leaving behind bound ONeFe WD remnants.
In that case, the final evolutionary products of NS+He star systems may be
the NS+WD binaries with eccentricity and relatively long orbital period
after the He stars undergo failed SN Ia explosions.
On the other hand,
long initial orbital periods or high kick velocities
may result in the disruption of the NS+WD binaries,
leaving isolated mildly recycled pulsars
\citep[][]{2017ApJ...846..170T}.

\subsection{Comparison to previous studies}\label{sec:comp}
For the evolution of single He star,
the amount of residual carbon in degenerate ONe core plays an important influence
on the triggering of explosive oxygen burning
\citep[][]{Antoniadis2020A&A}.
\citet{2022A&A...668A.106C} calculated a series of isolated He stars with different initial masses,
and found that at least $\sim0.003\,\rm M_\odot$ of residual carbon
is required to trigger explosive oxygen burning.
However,
the value of $M_{\rm c}(^{12}\rm C)$ calculated in our work is
about a few $10^{-4}\rm\,M_\odot$ (see Table\,\ref{table:1}),
which is smaller than the single He star channel.
This is because the initial single He star masses for SNe Ia are in the range of $\sim 2.0-2.5\rm\,M_\odot$,
which are lower than our results (see Table\,\ref{table:1}).
Compared to the single He star channel,
the He star companions in NS binaries can develop more massive CO cores before Case BB RLOF occurs
\citep[][]{2021A&A...656A..58L}.
Thus, the carbon burning in our simulations starts at the center rather than off-center,
thereby resulting in lower residual carbon abundance.
	In addition, we note that the residual carbon is gradually consumed during Ne-flash phase (see Fig.\,\ref{fig:ele-one}),
	indicating that the explosive oxygen burning is not caused by the residual carbon burning (see Fig.\,\ref{fig:pc-tc1-nm}).
	
\begin{figure}
	\centering\includegraphics[width=\columnwidth*5/5]{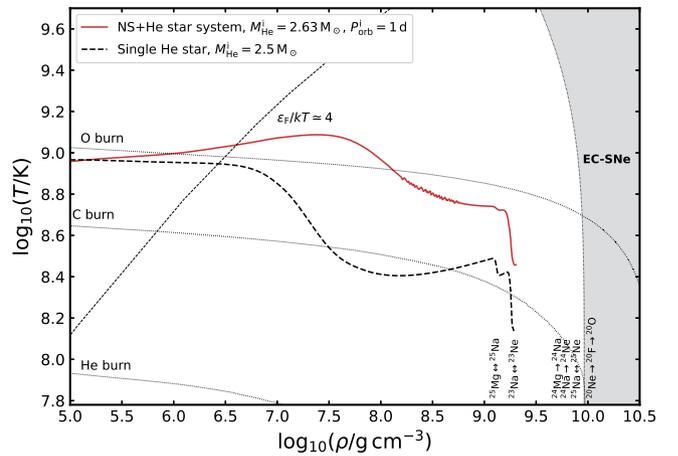}
	\caption{Profile of the density and temperature for two models after undergoing the Urca process.
		The red line represents the He star in NS binary,
		in which the initial mass of the He star is $2.63\rm\,M_\odot$
		and the initial orbital period is $1$\,d.
		The black dashed line represents the single He star with initial mass of $2.5\rm\,M_\odot$
		\citep[][]{Antoniadis2020A&A}.
	}
	\label{fig:pt-com}
\end{figure}

	Furthermore, we found that
	the mechanism of triggering oxygen ignition in this work is the convective Urca process
	\citep[e.g.][]{2005MNRAS.356..131L,2008NewAR..52..381P,2017MNRAS.472.3390S,2017ApJ...851..105S},
	which is different from the isolated He star channel.
Fig.\,\ref{fig:pt-com} show the $\rho-T$ profile of the single He star and the He star in NS binary
after undergoing the Urca process.
Compared with the single He star,
the He star in NS binary develops an ONe core with higher temperature.
During the Urca-process cooling phase,
the temperature gradient at the Urca reaction site gradually increase
and exceed adiabatic temperature gradient,
and thereby forming convection eventually. 

In the work of \citet{2022A&A...668A.106C},
the bottom of the convective envelope may touch He-burning shell (i.e. hot-bottom He-burning)
after an ONe core is formed,
resulting in a metal-rich envelope
and the significant expansion of the He star.
However,
for the NS+He star binaries in this work,
the stable mass-transfer process is maintained
after the onset of Case BB RLOF
(see Fig.\,\ref{fig:mass-loss}),
resulting in a rapid loss of He-rich envelope.
Thus,
the significant expansion of He star and the hot-bottom He-burning
do not occur in our simulations
\citep[e.g.][]{2013A&A...558A..39T, 2015MNRAS.451.2123T, 2018MNRAS.477..384L}.
Meanwhile,
the final metal core simulated by this work is surrounded
by a thin He-shell with a mass $\sim 3.6\times10^{-4}\,\rm M_\odot$ prior to SN.
\citet{2012MNRAS.422...70H} suggested that at least $0.06\rm\,M_\odot$ of remaining He envelope
is needed to observe the He lines in optical/IR spectra.
Thus, for the NS binaries with initial orbital periods $\lesssim10$\,d,
we do not expect to observe the He lines in the spectra after thermonuclear explosions.
For the isolated He star channel,
the envelope is metal-rich due to the hot-bottom He-burning prior to SN,
indicating that the He lines cannot be detected in the spectra.

Our simulations show that
the degenerate ONe cores could undergo Ne-flashes
and form ONeSi cores
if the core masses range from $\sim1.37-1.385\rm\,M_\odot$
\citep[e.g.][]{Nomoto1984, 2013ApJ...771...28T},
and then the ONeSi cores will explode as relatively low luminosity SNe Ia
due to the convective Urca process
(see Section \ref{sec:example}).
On the other hand,
the ONe cores will not experience the Ne-flashes
if the core masses are lower than $\sim 1.37\rm\,M_\odot$.
In this work,
if ONe cores with masses in the range of $\sim 1.335-1.37\rm\,M_\odot$ are formed
after He star companions experience central helium and carbon burning phase,
then the He star companions will undergo ONe core explosions caused by the convective Urca process.

Previous studies usually used a critical metal core mass of $\sim 1.37\,\rm M_\odot$
as a criterion for the production of EC-SNe
\citep[e.g.][]{1984ApJ...277..791N,2015MNRAS.451.2123T,2015MNRAS.446.2599D}.
\citet{2015MNRAS.451.2123T} performed a systematic study of the evolution of
NS+He star binaries with different initial He star masses ($2.5-3.5\,\rm M_\odot$)
and different initial orbital periods ($0.06-2$\,d),
and investigated the binary parameter space leading to
CO WDs, ONe WDs, EC-SNe and iron core-collapse SNe.
They used a critical core mass of $1.37\,\rm M_\odot$
as the boundary between ONe WDs and EC-SNe.
\citet{2015MNRAS.446.2599D} explored the final outcomes of intermediate-mass ZAMS stars,
and used $1.375\,\rm M_\odot$ as the critical mass for forming EC-SNe.
However, our simulations show that
if the metal core mass ranges from $1.335\,\rm M_\odot$ to $1.385\,\rm M_\odot$,
then the He stars may explode as SNe Ia because of the convective Urca process.
This may lead to a narrowing of the parameter space for EC-SNe.

\subsection{Other relevant discussions}
The delay times of SNe Ia are
the time interval between star formation and SN explosions.
Generally,
for primordial MS binaries to evolve into NS+He star systems
that can undergo SNe Ia,
the evolution time is about $25$\,Myr
\citep[e.g.][]{2006epbm.book.....E,2013ApJ...772..150J,2013ApJ...771...28T}.
Moreover,
for the He star companions with initial masses of $\sim2.5-2.8\,\rm M_\odot$,
the evolutionary time from He-ZAMS to explosive oxygen burning
is less than $5$\,Myr (see Table\,\ref{table:1}).
Accordingly, the delay times of SNe Ia from this channel is $\sim30$\,Myr,
which is shorter than that of the WD+He star channel
\citep[$\sim 100$\,Myr, see][]{2009MNRAS.395..847W}.
Thus, this work provides a channel for the formation of SNe Ia
with the shortest delay times so far.

For the classic SD model,
the progenitors of SNe Ia could appear as supersoft X-ray sources,
symbiotics and cataclysmic variables in observations
\citep[see, e.g.][]{2018RAA....18...49W}.
Compared with the SD model,
the progenitors of SNe Ia for the NS+He star channel show as strong X-ray sources
with luminosity $\rm \sim 10^{38}\,erg\,s^{-1}$,
or even ultraluminous X-ray binaries
if the Eddington limit is ignored
\citep[e.g.][]{2015ApJ...802..131S,2017ARA&A..55..303K}.
Moreover,
the NS+He star binaries with short orbital periods are
the potential GW sources in the low-frequency region
\citep[e.g.][]{2010NewAR..54...87N,2018PhRvL.121m1105T}. 

Aside from the contribution to the formation of SNe Ia,
NS+He star systems can also produce some other peculiar objects in the observations,
such as intermediate-mass binary pulsars, double NS binaries,
ultraluminous X-ray sources, millisecond pulsars (MSPs), etc
\citep[e.g.][]{2013MNRAS.432L..75C, 2013ApJ...778L..23T, 2015MNRAS.451.2123T, 2017ApJ...846..170T,
2019MNRAS.490..752T,2019ApJ...886..118S,2021ApJ...920L..36J}.
Especially,
\citet{2021MNRAS.506.4654W} recently suggested that NS+He systems can form
the ultracompact X-ray binaries (UCXBs) with relatively long orbital periods.
\citet{2022MNRAS.515.2725G} further considered evaporation process based on
the work of \citet{2021MNRAS.506.4654W},
and found that NS+He binaries can reproduce black widow pulsars with 
companion masses $\lesssim0.01\,\rm M_\odot$,
or even isolated MSPs within the Hubble time.

\section{Summary}
By using the stellar evolution code MESA,
we studied the formation of SNe Ia through the NS+He star channel
for the first time.
In our simulations,
the He star companions develop highly degenerate ONe cores
with masses around $M_{\rm Ch}$,
and then explosive oxygen burning can be triggered
owing to the convective Urca process.
We then performed a series of NS+He star systems with different
initial He star masses and orbital periods,
and obtained an initial parameter space for SNe Ia in the
$\rm log\,$$P^{\rm i}_{\rm orb}-M^{\rm i}_{\rm He}$ plane,
which can be used in the future binary population synthesis studies.
The results show that the final core masses for producing SNe Ia range from $\sim1.335\rm\,M_\odot$ to $1.385\rm\,M_\odot$.
He star companions may evolve into EC-SNe if the final core masses $\gtrsim1.385\rm\,M_\odot$
owing to $e$-capture on $\rm^{20}Ne$,
while the final fates of the He star companions are ONe WDs if the core masses $\lesssim1.335\rm\,M_\odot$.

In addition,
NSs can spin up by accreting the material from He star companions,
resulting in the formation of isolated mildly recycled pulsars
after the He stars explode as SNe Ia.
We found that for NS+He star binaries with $P^{\rm i}_{\rm orb} = 0.07-10$\,d,
such isolated pulsars produced by this channel have
minimum spin periods of $\sim 30-110$\,ms and
final orbital velocities of $\sim 60-360\,\rm km\,s^{-1}$.
By comparing our simulation results with the observations,
we speculate that this channel could produce
the isolated mildly recycled pulsars with
velocity $\rm \lesssim 360\,km\,s^{-1}$,
and such isolated pulsar are expected to locate in the region
of pulsars with massive WDs
in the $P_{\rm spin}-\dot P_{\rm spin}$ diagram.

\section*{Acknowledgements}
We acknowledge the anonymous referee for the valuable comments
that help to improve this paper.
We thank Professor Xiangcun Meng and Professor Zhengwei Liu
for useful discussions and comments.
This study is supported by the
the National Natural Science Foundation of China (Nos 12225304, 12288102, 12090040/12090043, 12003013, 12273014 and 11733009),
National Key R\&D Program of China (No. 2021YFA1600404),
the Western Light Project of CAS (No. XBZG-ZDSYS-202117),
the science research grant from the China Manned Space Project (No. CMS-CSST-2021-A12),
the Yunnan Fundamental Research Project (Nos 202001AS070029
and 202201BC070003),
the Frontier Scientific Research Program of Deep Space Exploration Laboratory (No. 2022-QYKYJH-ZYTS-016),
and the Shandong Fundamental Research Project (No. ZR2021MA013).
The authors also acknowledge the ``PHOENIX Supercomputing Platform''
jointly operated by the Binary Population Synthesis Group
and the Stellar Astrophysics Group at Yunnan Observatories, Chinese Academy of Sciences.
\section*{Data availability}
Results will be shared on reasonable request to corresponding author.

\bibliographystyle{mnras} 
\bibliography{1bib.bib}					



\label{lastpage}
\end{document}